\documentclass[showpacs, aps, prb, unsortedaddress]{revtex4-1}

\usepackage{amssymb}
\usepackage{latexsym}
\usepackage{amsmath}
\usepackage{graphicx}
\begin{document}

\title{Elastic Metamaterials and Dynamic Homogenization: A Review}

\date{\today}
\author{Ankit Srivastava}
\thanks{Corresponding Author}
\email{ankit.srivastava@iit.edu}

\affiliation{Department of Mechanical, Materials, and Aerospace Engineering
Illinois Institute of Technology, Chicago, IL, 60616
USA}

\begin{abstract}
In this paper we review the recent advances which have taken place in the understanding and applications of acoustic/elastic metamaterials. Metamaterials are artificially created composite materials which exhibit unusual properties which are not found in nature. We begin with presenting arguments from discrete systems which support the case for the existence of unusual material properties such as tensorial and/or negative density. The arguments are then extended to elastic continuums through coherent averaging principles. The resulting coupled and nonlocal homogenized relations, called the Willis relations, are presented as the natural description of inhomogeneous elastodynamics. They are specialized to Bloch waves propagating in periodic composites and we show that the Willis properties display the unusual behavior which is often required in metamaterial applications such as the Veselago lens. We finally present the recent advances in the area of transformation elastodynamics, charting its inspirations from transformation optics, clarifying its particular challenges, and identifying its connection with the constitutive relations of the Willis and the Cosserat types. 

\end{abstract}

\maketitle

\section{Introduction}
Metamaterials are artificially designed composite materials which can exhibit properties that can not be found in nature. These properties can be electronic, magnetic, acoustic, or elastic and have, of late, come to include static \cite{jang2013fabrication} properties as well. In the context of acoustic metamaterials these properties refer to the bulk modulus and density and for elastic metamaterials they refer to the moduli (bulk, shear, anisotropic) and density of a designed composite material. As such, they are used for the fine-tuned, predominantly frequency dependent control of the trajectory and dissipation characteristics of acoustic and stress waves. These materials have found natural applications in the research areas of cloaking, imaging, and noise and vibration control. The primary driver in acoustic metamaterials research has been research in the area of photonic metamaterials. As such, many of the conclusions drawn from the photonics research directly apply to acoustic waves and acoustic metamaterials due to the essential similarity of the governing equations in the two cases. Realizing analogous results for elastic metamaterials is complicated by the fact that the governing equation for elasticity admits both longitudinal and shear wave solutions which are capable of exchanging energy between each other. However, even in the case of elastic metamaterials some general ideas have been borrowed from photonic metamaterials research.

\begin{figure}[htp]
\centering
\includegraphics[scale=.5]{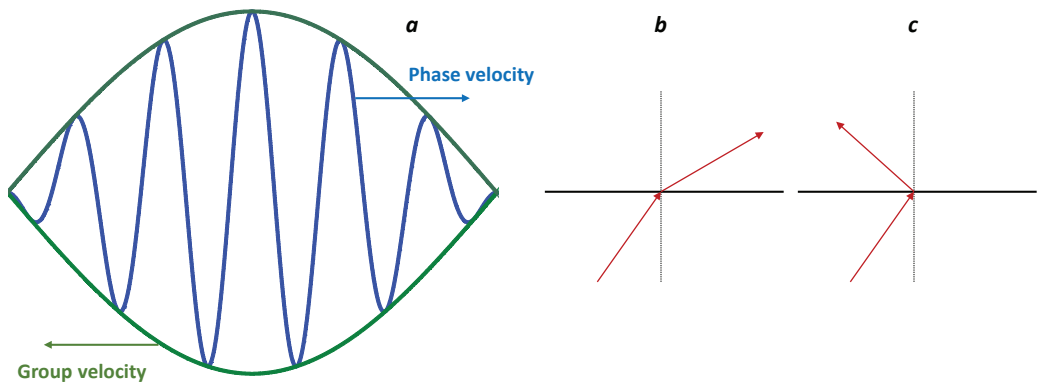}
\caption{Left handed materials and negative refraction, a. LHMs are characterized by anti-parallel phase and group velocity directions, b. refraction at an interface between two normal materials, c. refraction at an interface between a normal material and an LHM.}\label{LHM}
\end{figure}

There are two broad directions from which the research area of metamaterials can be approached. The first direction seeks to find the uses and applications of those materials which exhibit unnatural material properties such as negative density and moduli (negative $\epsilon$ and $\mu$ in the case of electromagnetism). In this approach, the question of the existence of such a material is secondary and the applications themselves are of primary importance. Within this approach researchers have conjectured that materials with simultaneously negative material properties will exhibit such exotic phenomena as negative refraction, reversed doppler effect, and reversed Cherenkov radiation \cite{veselago1968electrodynamics}. Such materials have been termed left handed materials (LHMs) and are characterized by the quality that a wave of appropriate frequency traveling through such a material will display anti-parallel phase and group velocity directions (Fig. \ref{LHM}). There are other scenarios, however, which allow for the existence of anti-parallel waves such as guided waves \cite{achenbach1984wave} and negative group velocity bands in photonic and phononic crystals (See also \cite{oliner1962backward,lindell2001bw,meisels2005refraction}). There exists the possibility of achieving negative refraction in such media as well \cite{notomi2000theory,sukhovich2008negative}. In fact, it has been shown recently that negative energy refraction can be accompanied by positive phase velocity refraction, and conversely that positive energy refraction can be accompanied by negative phase-velocity refraction \cite{nemat2014anti}. LHMs, on the other hand, can be uniquely attributed a negative index of refraction through arguments of causality \cite{smith2000negative}. This unique characteristic makes them suitable for applications in creating flat lenses which can beat the diffraction limit in imaging applications \cite{pendry2000negative}. If, in addition to LHMs, one could find materials with any desired material property then it becomes theoretically possible to design perfect cloaks which render an object enclosed within it completely invisible. The actual design of the cloak is approached through clever coordinate transformations which preseve the form of the governing equation making it possible to identify the desired change of wave trajectories with transformed material properties. This approach to designing cloaks has been applied to electromagnetic waves \cite{greenleaf2003anisotropic,greenleaf2003nonuniqueness,leonhardt2006notes,leonhardt2006optical,pendry2006controlling}, acoustic waves \cite{norris2008acoustic}, and elastic waves \cite{milton2006cloaking,norris2011elastic}.

The second broad direction in metamaterials research seeks to find those material microstructures which will display the unusual properties required for the application areas discussed earlier. No naturally occuring homogeneous material displays LHM properties of negative density and stiffness  \cite{li2004double}. Furthermore, naturally occuring materials offer only a very limited spectrum of density and moduli which is not enough to realize cloaking and other trajectory control applications. In order to achieve the properties required in the applications discussed earlier, researchers have sought to design heterogeneous materials at appropriate length scales which exhibit desirable effective properties \cite{smith2000composite}. 
\begin{figure}[htp]
\centering
\includegraphics[scale=.5]{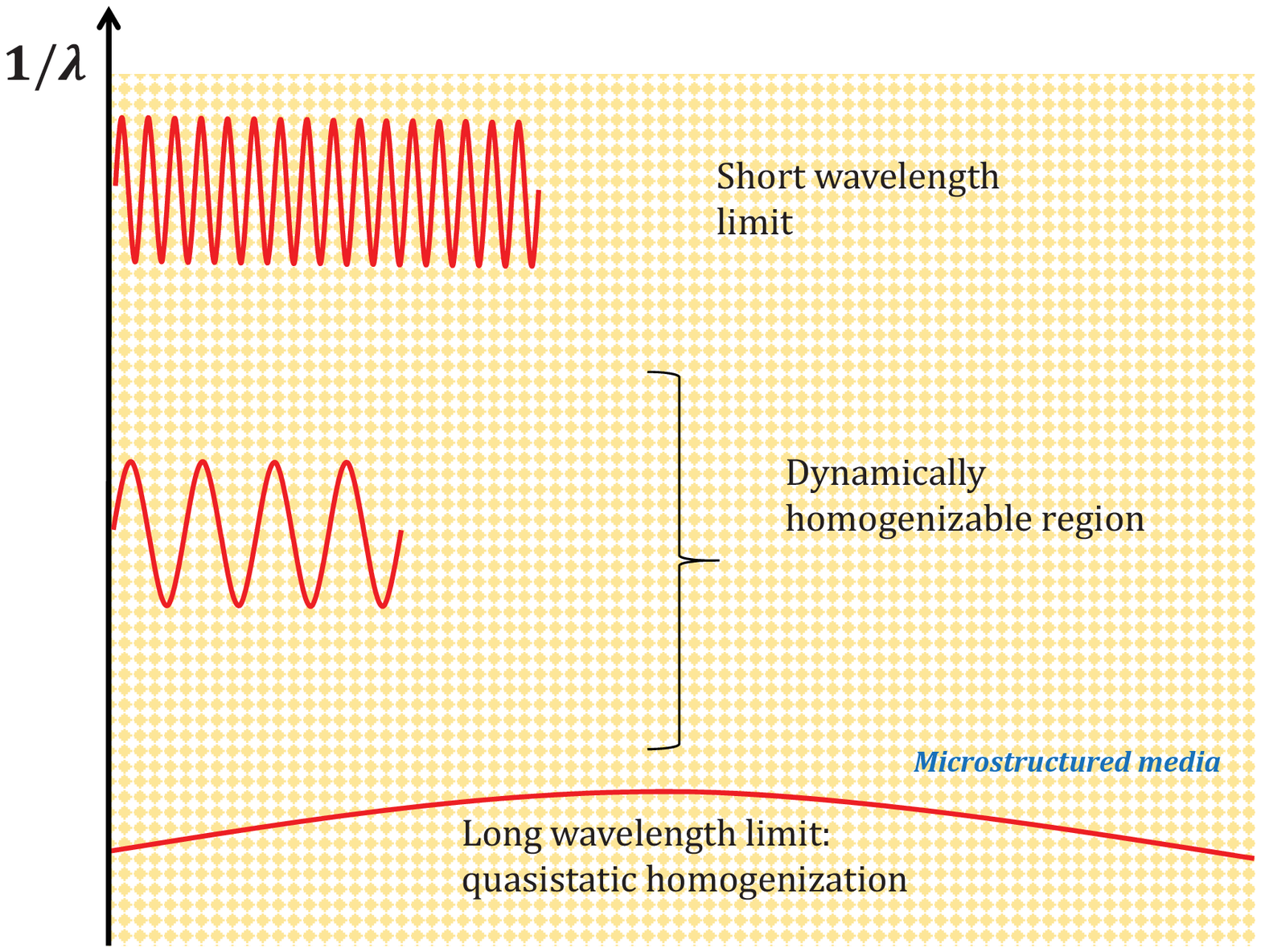}
\caption{Homogenizable region}\label{Hom}
\end{figure}

Fig. (\ref{Hom}) shows three broad regions in which wave phenomena can be studied. In the low-frequency region the predominant wavelength, $\lambda$, is much larger than the microstructural length scale of the material in which it is traveling. In this region wave characteristics are non-dispersive and propagation is controlled by the static averages of material properties. The traditional area of static homogenization is appropriate for determining density, $\rho(\mathbf{x})$, and stiffness tensor, $\mathbf{C}(\mathbf{x})$, which control wave behavior in this regime \cite{nemat1999micromechanics}. At the other end of the scale, the wavelength of the wave is on the same scale as or shorter than the length scale of the microstructure. Wave behavior in this regime is dominated by scattering at material interfaces and the heterogeneous material cannot be effectively defined by average homogenized properties. This is the case when one considers the optical branches in phononics or photonics \cite{kittel1976introduction}. Between these two scales lies a regime where the heterogeneous material may still be defined by homogenized effective properties but those properties must take into consideration the dispersive nature of wave propagation. This effectively means that in this regime the homogenized material properties which control wave propagation will need to be frequency dependent and, therefore, static homogenization techniques are not sufficient anymore. The primary problem, therefore, is one of relating the microstructure of a composite to the frequency dependent effective properties which will adequately represent wave phenomena in it and which are useful to the applications discussed above.

\section{Emergence of negative and tensorial material properties}

Metamaterial applications naturally require frequency dependence and additional tensorial complexity (e.g. tensorial density) of material properties. The concept of dispersive (frequency dependent) material properties naturally arises when laws of motion are enforced at scales below which additional heterogeneity, capable of dynamics, exist. Furthermore, this process of homogenization can give rise to tensorial forms of those material properties which are traditionally taken as scalar, such as density. 
\begin{figure}[htp]
\centering
\includegraphics[scale=.6]{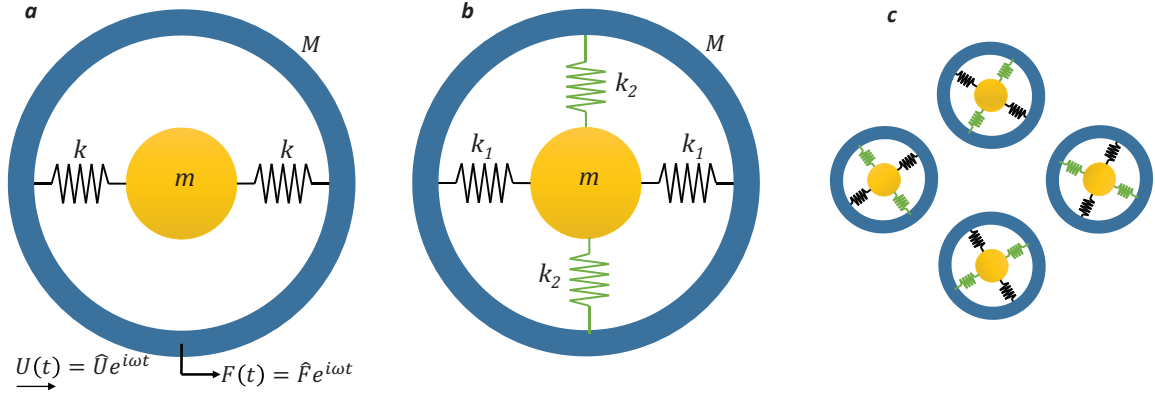}
\caption{Frequency dependent effective mass, a. 1-D, b,c. tensorial frequency dependent effective mass.}\label{ND}
\end{figure}
Consider, for instance, the dynamics of the one dimensional composite system shown in Fig. (\ref{ND}a) \cite{milton2007modifications,huang2009negative}. If we relate the harmonic macroscale force to the harmonic macroscale acceleration we find:
\begin{equation}\label{meff}
\hat{F}=-\omega^2M^\mathrm{eff}\hat{U};\quad M^\mathrm{eff}(\omega)=M+\frac{m\omega_0^2}{\omega_0^2-\omega^2};\quad \omega_0=\sqrt{2k/m}
\end{equation}
showing that by insisting on the applicability of Newton's second law at the macroscale, we inevitably end up with a frequency dependent effective mass. The appearance of mass dispersion directly results from the act of subsuming the microstructural dynamic effects into a homogenized macroscale description. $M^\mathrm{eff}$ increases to infinity as $\omega$ approaches $\omega_0$ (resonance condition) and becomes negative beyond that. By extending the idea to 2-, and 3-dimensions and by incorporating springs of different moduli, it is clear that the effective mass can not only be made frequency dependent but also anisotropic in nature. In 2-dimensions (Fig. \ref{ND}b), the relevant equations become:
\begin{equation}\label{meff3d}
\hat{F}_i=-\omega^2M^\mathrm{eff}_{ij}\hat{U}_j;\quad M^\mathrm{eff}_{ij}(\omega)=\left[M+\frac{m\omega_{[j]}^2}{\omega_{[j]}^2-\omega^2}\right]\delta_{ij};\quad \omega_{[j]}=\sqrt{2k_j/m};\quad i,j=1,2
\end{equation}
where $\delta_{ij}$ is the delta function. In the above, the off-diagonal terms of $\mathbf{M}^\mathrm{eff}$ are zero. They can be made non-zero, thereby coupling the displacements $U_j$, by incorporating the springs along axes which are not oriented along the axes of the orthogonal coordinate system \cite{milton2007modifications} (Fig. \ref{ND}c).

The assumption of harmonicity may seem restrictive in the above examples. In general, the time dependent macroscale displacement vector, $\mathbf{U}(t)$, would be related to the time dependent macroscale force vector, $\mathbf{F}(t)$, via a kernel, $\mathbf{H}(t)$, through the convolution operator, $\mathbf{F}=\mathbf{H}*\mathbf{U}$. However, it was shown by Milton et. al.  \cite{milton2007modifications} that the models of Fig. (\ref{ND}) are sufficiently rich to approximate the kernel under fairly unrestrictive conditions. By periodically repeating the microstructure, shrinking the unit cell size to zero while appropriately scaling the relevant quantities, one can obtain the justification for the existence of a frequency dependent mass density tensor $\boldsymbol{\rho}(\omega)$. Analogously to the frequency dependent mass of Eq. (\ref{meff}), the frequency dependent density is expected to be negative in some frequency range, the exact range depending upon the resonant behavior of the microstructure. If the same composite can also be made to exhibit negative modulus in the frequency range where its density is negative, then it would serve as an acoustic/elastic LHM with applicability to the metamaterial applications discussed above.

It must be mentioned that in the examples of Fig. (\ref{ND}) mass can only become negative in a homogenized effective sense and that a resonance is essential for achieving the negative behavior. However, effective mass can be tensorial and anisotropic in nature even away from the resonance (Eq. \ref{meff3d}). This requirement of resonance being important for achieving negative effective properties is seen in photonic metamaterials as well. A photonic LHM requires simultaneously negative dielectric permittivity ($\epsilon$) and magnetic permeability ($\mu$). Plasmon resonances in metals give rise to a frequency dependent dielectric permittivity which has a similar functional form as Eq. (\ref{meff}). Below the plasmon frequency $\epsilon$ is negative, however, at very low frequencies the effect of the plasmon is destroyed by dissipation. Pendry et. al. \cite{pendry1996extremely} showed that the plasmon frequencies can be reduced by introducing artificial resonances through a periodic assembly of thin metallic wires. It was further shown \cite{pendry1999magnetism} that a periodic array of nonmagnetic, resonant conducting units displays an effective $\mu$ which assumes negative values above the resonance. By combining the above, Smith et. al. \cite{smith2000composite} proposed a periodic array of interspaced conducting nonmagnetic split ring resonators and continuous wires which exhibited a frequency region where both $\epsilon^\mathrm{eff}$ and $\mu^\mathrm{eff}$ were simultaneously negative. 

The general ideas from photonic metamaterials research have inspired researchers to propose composite designs which use local mechanical resonances to exhibit negative effective density and moduli. In some cases researchers have even proposed negative designs which do not use resonances. The proposed designs began with Liu et. al. suggesting a design based on silicone coated lead spheres embedded in epoxy matrix \cite{liu2000locally,sheng2003locally}. It was subsequently suggested that acoustic/elastic LHMs could be constructed by mixing two structures which independently exhibit negative density and modulus \cite{ding2007metamaterial}. Negative modulus could be achieved through a periodic array of Helmholtz resonators \cite{fang2006ultrasonic,cheng2008one} the negative density could be achieved through an array of thin membranes \cite{lee2009acoustic}. The combination of these two structures was shown to exhibit LHM properties \cite{bongard2010acoustic,lee2010composite}. Since then, researchers have proposed more complicated acoustic/elastic LHM architectures. These include multiply resonant microstructures which lead to density and some components of the modulus tensor simultaneously becoming negative \cite{lai2011hybrid}, periodic array of curled 
perforations \cite{liang2012extreme,romero2013multi}, tunable piezoelectric resonator arrays \cite{casadei2012piezoelectric}, anisotropic LHMs by arranging layers of perforated plates \cite{christensen2012anisotropic} etc. It is clear that the materials which would satisfy the requirements for LHM properties will need to be dispersive since there doesn't exist any material with quasistatic LHM properties. This consequence places rather strong constraints on the broadband applicability of metamaterials to applications such as perfect cloaking and superlensing. Additionally, since LHM properties result from internal resonances, they are inevitably accompanied with large dissipation, thereby, limiting their use in practical applications. Some proposals have been made to compensate for the losses through active gain \cite{popov2006compensating}. However, there exist compelling theoretical arguments from causality which limit the potential benefit of such methods \cite{stockman2007criterion}.

\section{Dynamic homogenization}

The theoretical support to the field of metamaterials is provided through dynamic homogenization techniques which relate the microstructure of a composite to its frequency dependent dynamic effective properties. These effective properties 
must adequately represent the wave behavior in the composite within desired frequency ranges. The majority of research interest in the area of metamaterials is restricted to periodic composites which display highly dispersive wave behavior \cite{bloch1928quantum,john1987strong,yablonovitch1987inhibited,kushwaha1993acoustic,Sigalas1993,martinezsala1995sound}. These periodic composites admit Bloch waves as solutions and many different numerical algorithms have been developed for calculating the dispersive properties of these waves. These include the popular plane wave expansion method \cite{ho1990existence}, the finite difference time domain method \cite{chan1995order}, the multiple scattering method \cite{kafesaki1999multiple}, variational methods \cite{nemat1972harmonic,goffaux2003two,srivastava2014mixed}, secondary expansions \cite{hussein2009reduced} etc. 

\subsection{Averaging techniques}

There exist different ways by which effective constitutive parameters for wave propagation in metamaterials may be defined. The most common route to determining these parameters is by the use of retrieval methods \cite{o2002magnetic,koschny2003resonant,chen2004robust,smith2005electromagnetic} where the assumption is that local effective properties may be used to define periodic composites such as metamaterials. These properties are determined by analyzing the complex reflection and transmission coefficients of a finite slab of the composite. The retrieval method, which began with electromagnetic metamaterials, was subsequently extended to acoustic metamterials as well \cite{fokin2007method}. However, the retrieval method, while simple in principle, produces effective parameters which can fail to satisfy basic passivity and causality properties \cite{chaumet2004coupled} and exhibits other anti-resonant artifacts \cite{powell2010substrate,alu2011restoring,alu2011first}. An excellent review \cite{simovski2009material} points out that the majority of metamaterial homogenization studies published in the last decade failed to respect basic causality and passivity properties and in some cases also violated the second law of thermodynamics. Another broad technique for determining effective properties is through the Coherent Potential Approximation method \cite{shalaev1996electromagnetic} and its enhancements \cite{wu2006effective,hu2008homogenization}. The metamaterial under study is embedded in a matrix which has the properties of the effective media. These properties are then determined by minimizing wave scattering in the surrounding matrix and as such only apply in the long wavelength limit. Recent efforts which employ a similar idea of matching the surface responses of the structural unit of a metamaterial with a homogenized medium show applicability beyond the long wavelength limit \cite{yang2014homogenization}. The focus here is on the dynamic homogenization techniques which are geared towards extracting the effective dynamic properties of periodic acoustic/elastic composites which are applicable in the long wavelength limit and beyond. These properties are expected to represent Bloch wave propagation in such periodic composites and, as such, can be termed Bloch wave homogenization techniques. It must be noted that they are different from asymptotic homogenization \cite{bensoussan1978asymptotic,sanchez1980non,bakhvalov1989homogenisation} techniques which have traditionally been used for the calculation of the Bloch wave band-structure. Asymptotic homogenization was also limited in application to describing only the behavior of the fundamental Bloch mode at low frequencies \cite{parnell2006dynamic,andrianov2008higher}. Recently, however, it has been extended to higher frequencies and higher Bloch modes as well \cite{craster2009mechanism,craster2010high,antonakakis2013asymptotics,antonakakis2014homogenisation} (See also \cite{nagai2004stabilized}). Additionally, there have been other efforts to bridge the scales for the study of dispersive systems based upon variational formulations \cite{mcdevitt2001assumed}, micromechanical techniques \cite{wang2002modeling}, Fourier transform of the elastodynamic equations \cite{gonella2008homogenization}, and strain projection methods \cite{hussein2006mode}.

\subsection{Ensemble averaging}

\begin{figure}[htp]
\centering
\includegraphics[scale=.5]{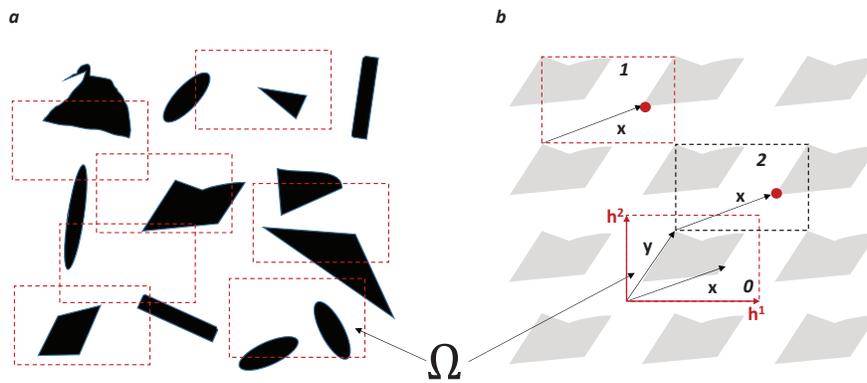}
\caption{Ensemble averaging, a. random composite, b. periodic composite.}\label{ENS}
\end{figure}

The pioneering work in the area of homogenization of inhomogeneous electromagnetics/elastodynamics was done by Willis \cite{willis1981avariational,willis1981bvariational,willis1983overall,willis1984variational}. Here we present the basic ideas which have led to the current state of the art in the area of dynamic homogenization. We consider a volume $\Omega$ in which the equations of motion and kinematic relations are given pointwise by:
\begin{equation}\label{equationofmotion}
\sigma_{ij,j}+f_i=\dot{p}_i; \quad \varepsilon_{ij}=\frac{1}{2}(u_{i,j}+u_{j,i}),
\end{equation}
where $\boldsymbol{\sigma},\mathbf{f},\boldsymbol{\varepsilon},\mathbf{p}$, and $ \mathbf{u}$ are the space and time dependent stress tensor, body force vector, strain tensor, momentum vector, and displacement vector, respectively. The pointwise constitutive relations are:
\begin{equation}\label{constitutive}
\sigma_{ij}=C_{ijkl}\varepsilon_{kl}; \quad p_i=\rho\dot{u}_i,
\end{equation}
with the usual symmetries for $\mathbf{C}$. Eqs. (\ref{equationofmotion},\ref{constitutive}) are supplemented with appropriate boundary conditions on $\partial\Omega$ and initial conditions at $t=0$. We seek to average the equations of motion and find the constitutive parameters which relate the averaged field variables $\langle\boldsymbol{\sigma}\rangle$, $\langle\boldsymbol{\varepsilon}\rangle$, $\langle\mathbf{p}\rangle$, and $\langle\dot{\mathbf{u}}\rangle$. The homogenization procedure can be derived for random composites and subsequently specialized to the periodic case \cite{willis1997dynamics}. Random composites represent families whose physical properties vary not only with position $\mathbf{x}$ but also with a parameter $\alpha$. $\alpha$ is a member of a sample space, $\mathcal{A}$, over which a probability measure $\mathcal{P}$ is defined. In essence, we solve many problems over $\Omega$ where $\Omega$ is defined, for each problem, by materials with properties $\mathbf{C}(\mathbf{x},\alpha),\rho(\mathbf{x},\alpha)$ (Fig. \ref{ENS}a). Each problem has the same body force, initial conditions, and boundary conditions. This leads to the pointwise solution $\mathbf{u}(\mathbf{x},\alpha)$ (and other derived field variables) for each problem. Now ensemble averages are defined over the sample space:
\begin{equation}\label{ensemble}
\langle\phi\rangle(\mathbf{x})=\int_\mathcal{A}\phi(\mathbf{x},\alpha)\mathcal{P}(d\alpha)
\end{equation}
Ensemble averaging the equation of motion (\ref{equationofmotion}) we get:
\begin{equation}\label{equationofmotionE}
\langle\sigma\rangle_{ij,j}+f_i=\langle\dot{p}\rangle_i;\quad \langle\varepsilon\rangle_{ij}=\frac{1}{2}(\langle u\rangle_{i,j}+\langle u\rangle_{j,i})
\end{equation}
which could be solved if the ensemble averages of stress and momentum could be related to the ensemble averages of strain and velocity through appropriate homogenized relations. It should be noted that such relations cannot be directly derived from averaging Eq. (\ref{constitutive}).

\subsection{Homogenized properties}

At this point a comparison medium is introduced with homogeneous properties ($\mathbf{C}^0,\rho^0$) which transforms Eq. (\ref{constitutive}) to:
\begin{equation}\label{constitutiveCom}
\sigma_{ij}=C_{ijkl}^0\varepsilon_{kl}+\Sigma_{ij}; \quad p_i=\rho^0\dot{u}_i+P_i,
\end{equation} 
where $\boldsymbol{\Sigma},\mathbf{P}$ are stress and momentum polarizations (See \cite{hashin1959moduli,hashin1962some,hashin1962variational,hashin1963theory} for polarizations). The above could be averaged and the required homogenized relation extracted if the ensemble averages of $\boldsymbol{\Sigma},\mathbf{P}$ could be determined. Using Eq. (\ref{constitutiveCom}) in the equation of motion we have:
\begin{equation}\label{equationofmotionCom}
(C_{ijkl}^0u_{k,l})_{,j}+f_i+\Sigma_{ij,j}-\dot{P}_i=\rho^0\ddot{u}_i
\end{equation}
Taking into consideration the initial and boundary conditions, the solution to Eq. (\ref{equationofmotionCom}) can be written as:
\begin{equation}\label{solution}
u_i(\mathbf{x},t)=u_i^0(\mathbf{x},t)+\int_t\int_\Omega G_{ij}^0(\mathbf{x},t;\mathbf{x}',t')\left[\Sigma_{jk,k}(\mathbf{x}',t')-\dot{P}_j(\mathbf{x}',t')\right]d\mathbf{x}'dt'
\end{equation}
where $\mathbf{u}^0$ is the solution to 
\begin{equation}\label{comparison}
(C_{ijkl}^0u_{k,l})_{,j}+f_i=\rho^0\ddot{u}_i
\end{equation}
with the same boundary conditions as in (\ref{equationofmotionCom}) and $\mathbf{G}^0$ is the Green's function of the comparison medium satisfying:
 \begin{equation}\label{green}
C^0_{jikl}G^0_{kp,jl}+\delta_{ip}\delta(\mathbf{x}-\mathbf{x}')\delta(t-t')=\rho^0\delta_{ik}\ddot{G}^0_{kp}
\end{equation}
with appropriate homogeneous boundary conditions on $\partial\Omega$.  Integration by parts of (\ref{solution}) formally gives:
\begin{equation}\label{solutionF}
\mathbf{u}=\mathbf{u}^0-\mathbf{S}^0\otimes\boldsymbol{\Sigma}-\mathbf{M}^0\otimes\mathbf{P}
\end{equation}
where $\otimes$ represents convolution in space and time as shown in Eq. (\ref{solution}). After appropriate space and time differentials we have:
\begin{eqnarray}
\displaystyle \nonumber \boldsymbol{\varepsilon}=\boldsymbol{\varepsilon}^0-\mathbf{S}^{0}_x\otimes\boldsymbol{\Sigma}-\mathbf{M}^{0}_x\otimes\mathbf{P}\\
\dot{\mathbf{u}}=\dot{\mathbf{u}}^0-\mathbf{S}^{0}_t\otimes\boldsymbol{\Sigma}-\mathbf{M}^{0}_t\otimes\mathbf{P}
\end{eqnarray}
where $\mathbf{S}^0_x,\mathbf{M}^0_x,\mathbf{S}^0_t,\mathbf{M}^0_t$ are integral operators \cite{willis1997dynamics}. The above can be used to express $\boldsymbol{\Sigma},\mathbf{P}$ in terms of $\langle\boldsymbol{\varepsilon}\rangle,\langle\dot{\mathbf{u}}\rangle$ after eliminating $\boldsymbol{\varepsilon}^0,\dot{\mathbf{u}}^0$ (See \cite{milton2007modifications} for details). This relation is formally given:
\begin{eqnarray}
\displaystyle \nonumber \boldsymbol{\Sigma}=\mathbf{T}_{11}\otimes\langle\boldsymbol{\varepsilon}\rangle+\mathbf{T}_{12}\otimes\langle\dot{\mathbf{u}}\rangle\\
\mathbf{P}=\mathbf{T}_{21}\otimes\langle\boldsymbol{\varepsilon}\rangle+\mathbf{T}_{22}\otimes\langle\dot{\mathbf{u}}\rangle
\end{eqnarray}
It is clear from the above that the ensemble averages of $\boldsymbol{\Sigma},\mathbf{P}$ depend on both $\langle\boldsymbol{\varepsilon}\rangle$ and $\langle\dot{\mathbf{u}}\rangle$. In conjunction with the ensemble average of Eq. (\ref{constitutiveCom}), it means that $\langle\boldsymbol{\sigma}\rangle$ and $\langle\mathbf{p}\rangle$ will depend upon both $\langle\boldsymbol{\varepsilon}\rangle$ and $\langle\dot{\mathbf{u}}\rangle$. The coupled homogenized constitutive relations which naturally emerge from ensemble averaging are formally given as:
\begin{eqnarray}\label{homogenized}
\displaystyle \nonumber \langle\boldsymbol{\sigma}\rangle=\mathbf{C}^\mathrm{eff}\otimes\langle\boldsymbol{\varepsilon}\rangle+\mathbf{S}^\mathrm{eff}\otimes\langle\dot{\mathbf{u}}\rangle\\
\langle\mathbf{p}\rangle=\bar{\mathbf{S}}^\mathrm{eff}\otimes\langle\boldsymbol{\varepsilon}\rangle+\boldsymbol{\rho}^\mathrm{eff}\otimes\langle\dot{\mathbf{u}}\rangle
\end{eqnarray}

A material which exhibits the above coupled constitutive relation, which is a generalization of the classical elastic constitutive relation, will be termed a Willis material. It is interesting to note from Norris et. al. \cite{norris2012analytical} that a periodic composite formed using Willis materials, under the dynamic homogenization process shown in this section, results in an effective dynamic constitutive relation which is again of the Willis kind. This shows that the Willis constitutive relation given above is closed under homogenization whereas the classical constitutive relation is not. 

\subsection{Specialization to Bloch/Floquet waves in periodic composites}\label{Bloch}

It should be noted that the averaged fields in Eq. (\ref{homogenized}) depend upon $\mathbf{x},t$ and that the constitutive properties appearing in (\ref{homogenized}) are integral operators in both space and time domains. These operators are considerably simplified when applied to the case of Bloch waves in periodic composites. Ensemble averages can be specialized to the periodic case. In Fig. (\ref{ENS}b) the unit cell denoted by index 0, $\Omega$, is characterized by base vectors $\mathbf{h}^i$. The reciprocal base vectors of the unit cell are given by,
\begin{equation}
\mathbf{q}^1=2\pi\frac{\mathbf{h}^2\times\mathbf{h}^3}{\mathbf{h}^1\cdot(\mathbf{h}^2\times\mathbf{h}^3)};\quad \mathbf{q}^2=2\pi\frac{\mathbf{h}^3\times\mathbf{h}^1}{\mathbf{h}^2\cdot(\mathbf{h}^3\times\mathbf{h}^1)};\quad \mathbf{q}^3=2\pi\frac{\mathbf{h}^1\times\mathbf{h}^2}{\mathbf{h}^3\cdot(\mathbf{h}^1\times\mathbf{h}^2)}
\end{equation}
such that $\mathbf{q}^i\cdot\mathbf{h}^j=2\pi\delta_{ij}$. The rest of the composite can be generated by repeating $\Omega$ such that the material properties have the following periodicity:
\begin{equation}
C_{jkmn}(\mathbf{x}+n_i\mathbf{h}^i)=C_{jkmn}(\mathbf{x});\quad \rho(\mathbf{x}+n_i\mathbf{h}^i)=\rho(\mathbf{x}),
\end{equation}
where $n_i$ are integers. The infinite periodic composite, thus generated, accepts Bloch waves as solutions with a wave vector specified by $\mathbf{k}=Q_i\mathbf{q}^i$. Any field variable $\mathbf{\Phi}(\mathbf{x},t)$ (stress, strain, velocity, or momentum) can be expressed as $\hat{\mathbf{\Phi}}(\mathbf{x})\exp\left[i(\mathbf{k}\cdot\mathbf{x}-\omega t)\right]$ where $\hat{\mathbf{\Phi}}$ is $\Omega$ periodic. Now consider another realization of the composite generated by repeating $\Omega^{(1)}$ (Fig. \ref{ENS}b). $\Omega^{(1)}$ is chosen randomly but due to the periodicity of the composite it is equivalent to some $\Omega^{(2)}$ which is translated from $\Omega$ by a vector $\mathbf{y}$ such that $\mathbf{y}\in\Omega$. This realization also accepts Bloch wave solutions of the form $\hat{\mathbf{\Phi}}^y(\mathbf{x})\exp\left[i(\mathbf{k}\cdot\mathbf{x}-\omega t)\right]$ where the superscript $y$ denotes the translated unit cell and $\hat{\mathbf{\Phi}}^y(\mathbf{x})$ is again $\Omega$ periodic. As can be seen from Fig. (\ref{ENS}b) any $\Omega$ periodic quantity $\hat{\mathbf{\Phi}}^y(\mathbf{x})$ is equivalent to $\hat{\mathbf{\Phi}}(\mathbf{x+y})$. The medium translated by $\mathbf{y}$ can be regarded as one of a statistical ensemble if $\mathbf{y}$ is assumed to be uniformly distributed over $\Omega$. Ensemble averaging of the field variables is, therefore, equivalent to:
\begin{equation}\label{ensembleP}
\langle\mathbf{\Phi}\rangle(\mathbf{x})=\int_\Omega\hat{\mathbf{\Phi}}^y(\mathbf{x})e^{i(\mathbf{k}\cdot\mathbf{x}-\omega t)}d\mathbf{y}=\left[\int_\Omega\hat{\mathbf{\Phi}}(\mathbf{x+y})d\mathbf{y}\right]e^{i(\mathbf{k}\cdot\mathbf{x}-\omega t)}=\langle\hat{\mathbf{\Phi}}\rangle e^{i(\mathbf{k}\cdot\mathbf{x}-\omega t)}
\end{equation}
which is the usual unit cell averaging \emph{carried over the $\Omega$ periodic parts of the field variables and not the full field variables}. It can be shown that when the averaging is performed over the $\Omega$ periodic parts of the field variables then the resulting effective properties automatically satisfy the dispersion relation of the periodic composite \cite{nemat2011homogenization}. For periodic composites the homogenized effective constitutive relation (\ref{homogenized}) can be written in the tensorial form \cite{srivastava2012overall}:
\begin{eqnarray}\label{homogenizedB}
\displaystyle \nonumber \langle{\hat{\sigma}}\rangle_{ij}={C}^\mathrm{eff}_{ijkl}\langle{\hat{\varepsilon}}\rangle_{kl}+{S}^\mathrm{eff}_{ijk}\langle\hat{\dot{{u}}}\rangle_{k}\\
\langle\hat{{p}}\rangle_{i}=\bar{{S}}^\mathrm{eff}_{ijk}\langle{\hat{\varepsilon}}\rangle_{jk}+{\rho}^\mathrm{eff}_{ij}\langle\hat{\dot{{u}}}\rangle_{j}
\end{eqnarray}
where $\mathbf{C}^\mathrm{eff},\mathbf{S}^\mathrm{eff},\bar{\mathbf{S}}^\mathrm{eff},\boldsymbol{\rho}^\mathrm{eff}$ are functions of $\mathbf{k},\omega$. Furthermore, the constitutive tensors display the following additional symmetries: 
\begin{eqnarray}\label{symm}
\displaystyle \nonumber C^\mathrm{eff}_{ijkl}=C^\mathrm{eff}_{jikl}=C^\mathrm{eff}_{ijlk}=(C^\mathrm{eff}_{klij})^*\\
\displaystyle \nonumber \bar{S}^\mathrm{eff}_{ijk}=({S}^\mathrm{eff}_{jki})^*\\
\displaystyle \rho^\mathrm{eff}_{ij}=(\rho^\mathrm{eff}_{ji})^*
\end{eqnarray}
where $*$ represents a complex conjugate. The above relations hold generally for Bloch waves in 3-dimensional linear periodic composites and are in congruence with effective dynamic properties for electromagnetic Bloch/Floquet waves \cite{amirkhizi2008microstructurally}. In one dimension these relations simplify further \cite{nemat2011overall}:
\begin{eqnarray}\label{homogenizedB1}
\displaystyle \nonumber \langle{\hat{\sigma}}\rangle=C^\mathrm{eff}\langle{\hat{\varepsilon}}\rangle+S^\mathrm{eff}\langle\hat{\dot{{u}}}\rangle\\
\langle\hat{{p}}\rangle=\bar{S}^\mathrm{eff}\langle{\hat{\varepsilon}}\rangle+\rho^\mathrm{eff}\langle\hat{\dot{{u}}}\rangle
\end{eqnarray}
where it can be shown that $C^\mathrm{eff}$ and $\rho^\mathrm{eff}$ are real and $\bar{S}^\mathrm{eff}=(S^\mathrm{eff})^*$. Moreover, $\bar{S}^\mathrm{eff},S^\mathrm{eff}\rightarrow 0$ as $\omega\rightarrow 0$ whereas $C^\mathrm{eff}$ and $\rho^\mathrm{eff}$ approach their quasistatic homogenized limits as $\omega\rightarrow 0$. Explicit calculations of effective properties in 3- and 1-D periodic composites are provided in Refs. \cite{srivastava2012overall,norris2012analytical,nemat2011overall}.

\subsection{Non-uniqueness of the homogenized relations and LHM properties}

Eq. (\ref{homogenizedB1}) are homogenized effective dynamic constitutive relations \emph{of one kind} for Bloch wave propagation in 1-D periodic composites. It will be shown later that this form of the constitutive relations may be important for acoustic/elastic cloaking applications. However, these relations are not unique and can be transformed into a form which is more directly applicable to left handed materials. The nonuniqueness of the effective relations results directly from the fact that they involve integral operators in the space and time domains of fields which are derived essentially from the same displacement field (\ref{homogenized}) . It is possible to transform $\mathbf{C}^\mathrm{eff}$ to $\mathbf{C}^\mathrm{eff}+\hat{\mathbf{C}}^\mathrm{eff}$ and $\mathbf{S}^\mathrm{eff}$ to $\mathbf{S}^\mathrm{eff}+\hat{\mathbf{S}}^\mathrm{eff}$ with appropriate conditions on $\hat{\mathbf{C}}^\mathrm{eff},\hat{\mathbf{S}}^\mathrm{eff}$ such that Eq. ($15^a$) is preserved. Similarly ($15^b$) could be preserved under appropriate transformations to $\bar{\mathbf{S}}^\mathrm{eff},\boldsymbol{\rho}^\mathrm{eff}$. These conditions are described in Ref. \cite{willis2011effective} where it is also pointed out that the nonuniqueness disappears if one assumes the existence of an inelastic strain in the body. For the electromagnetic case a similar result was proven by Fietz and Shvets \cite{fietz2009metamaterial} where they showed that the analogous electromagnetic effective dynamic constitutive relation can be determined uniquely only in the presence of a magnetic monopole current. In the absence of inelastic strain, there exists considerable freedom in the definitions of the effective parameters(See \cite{shuvalov2011effective} for instance).

As a consequence of the nonuniqueness of the effective dynamic constitutive relations, Eq. (\ref{homogenizedB1}), in 1-D, can be transformed so as to subsume the effects of $S^\mathrm{eff},\bar{S}^\mathrm{eff}$ into a modified set of effective modulus and density parameters \cite{willis2009exact,nemat2011homogenization}:
\begin{equation}\label{homogenizedB1m}
\langle\hat{\sigma}\rangle=\bar{C}^\mathrm{eff}\langle\hat{\varepsilon}\rangle;\quad \langle\hat{p}\rangle=\bar{\rho}^\mathrm{eff}\langle\hat{\dot{u}}\rangle
\end{equation}
The modified parameters $\bar{C}^\mathrm{eff},\bar{\rho}^\mathrm{eff}$ are functions of frequency and automatically satisfy the dispersion relation of the composite:
\begin{equation}\label{disp}
\sqrt{\frac{\bar{C}^\mathrm{eff}}{\bar{\rho}^\mathrm{eff}}}=\frac{\omega}{k}
\end{equation}
where $k$ is the one dimensional wavenumber.
\begin{figure}[htp]
\centering
\includegraphics[scale=.7]{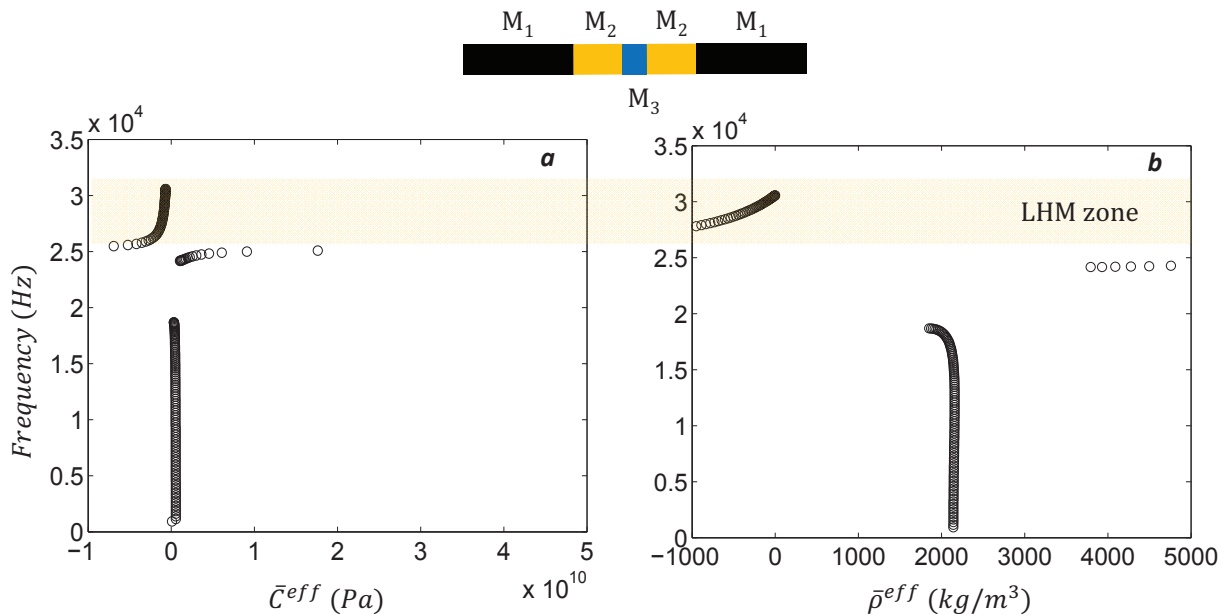}
\caption{Effective properties for an internally resonant 1-D unit cell, a. $\bar{C}^\mathrm{eff}$, b. $\bar{\rho}^\mathrm{eff}$}\label{LHM3p}
\end{figure}

Fig. (\ref{LHM3p}) shows effective properties calculated for a 3 phase composite where the central phase ($M_3$) is stiff and heavy and can resonate due to the light and compliant $M_2$ phase. This unit cell is the one dimensional equivalent of the locally resonant structure considered by Liu et. al. \cite{liu2000locally}. The geometrical and material properties of the unit cell are provided in Ref. \cite{nemat2011negative}. These properties are calculated from Eq. (\ref{homogenizedB1m}) over the first two branches of the phononic bandstructure of the composite. It is clear that for this unit cell both $\bar{C}^\mathrm{eff},\bar{\rho}^\mathrm{eff}$ are simultaneously negative in the frequency region denoted by the yellow rectangle in Fig. (\ref{LHM3p}). This frequency region corresponds to the Veselago left hand metamaterial zone (See also \cite{liu2011elastic}).

\subsection{Applicability of the homogenized relations}

It should be noted that Eq. (\ref{homogenized}) is exact for a given elastodynamic problem on $\Omega$ if the boundary conditions of the problem are appropriately represented in the associated homogeneous problem (\ref{comparison}). However, this is often very complicated and the effort is to represent finite composites with homogenized properties which correspond to an infinite body. This is a valid approach in the homogenization limit \cite{kohn2008magnetism} and is equivalent to ascribing the Bloch homogenized properties presented in section (\ref{Bloch}) to the finite and or semi-infinite cases which appear in focusing, negative refraction, and cloaking applications. Refs. \cite{shuvalov2011effective,srivastava2014limit,willis2013some} have studied this approximation and concluded that the approximation grows worse as frequency is increased. However, the approximation of homogenization for 3-phase locally resonant unit cells is, in general, better than it is for non-resonant unit cell. Willis \cite{willis2013some} has further studied the applicability of a modified form of the homogenized relations based upon weighted averages of the fields.

\section{Coordinate transformations and metamaterials}

In general, research in electromagnetic/acoustic/elastic wave cloaking has developed independently of the research in metamaterials/dynamic homogenization but they have converged to some common ideas and are intricately related now. Research in electromagnetic cloaking began with Pendry's observation that Maxwell's equations preserve their form under coordinate transformations, albeit with modified $\epsilon,\mu$ \cite{post1997formal,ward1996refraction,ward1998program}. Specifically, Maxwell's equations at a fixed frequency, $\omega$, are:
\begin{eqnarray}\label{max}
\displaystyle \nonumber \nabla\times\mathbf{E}+i\omega\boldsymbol{\mu}\mathbf{H}=0;\quad \nabla\times\mathbf{H}-i\omega\boldsymbol{\epsilon}\mathbf{E}=0
\end{eqnarray}
under the coordinate transformation $\mathbf{x}'=\mathbf{x}'(\mathbf{x})$, they retain their form:
\begin{eqnarray}\label{maxt}
\displaystyle \nonumber \nabla '\times\mathbf{E}'+i\omega\boldsymbol{\mu}'\mathbf{H}'=0;\quad \nabla '\times\mathbf{H}'-i\omega\boldsymbol{\epsilon}'\mathbf{E}'=0
\end{eqnarray}
with the following transformed properties:
\begin{eqnarray}\label{maxtp}
\displaystyle \nonumber \boldsymbol{\mu}'(\mathbf{x}')=\mathbf{A}\boldsymbol{\mu}(\mathbf{x})\mathbf{A}^\mathrm{T}/det\mathbf{A};\quad \boldsymbol{\epsilon}'(\mathbf{x}')=\mathbf{A}\boldsymbol{\epsilon}(\mathbf{x})\mathbf{A}^\mathrm{T}/det\mathbf{A}
\end{eqnarray}
where $A_{ij}=\partial x^{'}_{i}/\partial x_{j}$ (and transformed fields as well). This observation was used to design a cloak through coordinate transformations which mapped the trajectories of electromagnetic waves in the presence of a cavity surrounded by a cloak to a homogeneous material \cite{leonhardt2006optical,pendry2006controlling}. This ensured that the cavity would become invisible to light waves generated by any source and incident at it from any direction. There has been considerable research activity in the field since then (See \cite{chen2010transformation} for a review). 

Cummer and Schurig \cite{cummer2007one} showed that there exists complete isomorphism between electromagnetic waves and acoustic waves in 2-D. In cylindrical coordinates with $z$ invariance and accounting for a tensorial but diagonal mass density, the acoustic wave equations and constitutive relation are given by:
\begin{eqnarray}\label{acoustic}
\displaystyle \nonumber i\omega\rho_\phi v_\phi=-\frac{1}{r}\frac{\partial p}{\partial \phi};\quad i\omega\rho_rv_r=-\frac{\partial p}{\partial r}\\
i\omega p=-\frac{\lambda}{r}\left[\frac{\partial(rv_r)}{\partial r}+\frac{\partial v_\phi}{\partial \phi}\right]
\end{eqnarray}
where $\lambda$ is the bulk modulus of the fluid, $p$ is pressure, and $\mathbf{v}$ is vector fluid velocity. These compare with z-invariant Maxwell's equations in cylindrical coordinates for TE polarization:
\begin{eqnarray}\label{maxwell}
\displaystyle \nonumber i\omega\mu_r (-H_r)=-\frac{1}{r}\frac{\partial (-E_z)}{\partial \phi};\quad i\omega\mu_\phi H_\phi=-\frac{\partial (-E_z)}{\partial r}\\
i\omega (-E_z)=-\frac{1}{\epsilon_zr}\left[\frac{\partial(rH_\phi)}{\partial r}+\frac{\partial (-H_r)}{\partial \phi}\right]
\end{eqnarray}
$\left[p,v_r,v_\phi,\rho_r,\rho_\phi,\lambda^{-1}\right]\rightleftharpoons$ $\left[-E_z,H_\phi,-H_r,\mu_\phi,\mu_r,\epsilon_z\right]$ clarifies the duality between Eqs. (\ref{acoustic},\ref{maxwell}). Once the duality is established, designs for electromagnetic cloaks can be easily transformed to designs for acoustic cloaks. In the 3-D case Chen and Chan \cite{chen2007acoustic} noted that the acoustic wave equations retain their form but with transformed material properties. Specifically, the acoustic wave equation at a fixed frequency, $\omega$, is:
\begin{eqnarray}\label{ac}
\displaystyle \nonumber \nabla\left[\frac{1}{\rho(\mathbf{x})}\nabla p(\mathbf{x})\right]=-\frac{\omega^2}{{\kappa}(\mathbf{x})}p(\mathbf{x})
\end{eqnarray}
where $p$ is the fluid pressure and $\kappa$ is the bulk modulus of the fluid. Under the coordinate transformation $\mathbf{x}'=\mathbf{x}'(\mathbf{x})$, the acoustic wave equation retains its form:
\begin{eqnarray}\label{act}
\displaystyle \nonumber \nabla '\left[\frac{1}{\boldsymbol{\rho}'(\mathbf{x})}\nabla ' p'(\mathbf{x}')\right]=-\frac{\omega^2}{{\kappa}'(\mathbf{x}')}p'(\mathbf{x}')
\end{eqnarray}
with the following transformed properties:
\begin{eqnarray}\label{actp}
\displaystyle \nonumber 
\frac{1}{\boldsymbol{\rho}'(\mathbf{x}')}=\mathbf{A}\frac{1}{{\rho}(\mathbf{x})}\mathbf{A}^\mathrm{T}/det\mathbf{A}; {\kappa}'(\mathbf{x}')=\kappa(x)det\mathbf{A}
\end{eqnarray}
Once the invariance of the acoustic equation was noted, ideas from electromagnetic cloak design were brought to bear upon the design of acoustic cloaks. The anisotropic densities needed to realize such cloaks could be achieved by using layered fluids \cite{torrent2008acoustic,norris2009acoustic} (See also \cite{mei2007effective,torrent2008anisotropic}), however Norris \cite{norris2008acoustic} realized that such \emph{inertial} cloaks suffer from a considerable mass penalty. He presented an alternative route to designing cloaks for acoustic waves based upon  pentamode materials \cite{milton1995elasticity,norris2008acoustic,gokhale2012special}. The existence of different routes to acoustic wave cloaking has been noted in literature \cite{hu2013constraint}. For both electromagnetic and acoustic cloaking cases, the problem boils down to finding the microstructures which would exhibit those effective properties which are required by an appropriate coordinate transformation. An excellent review exists on the topic of transformational acoustics and its application to the design of acoustic cloaks \cite{chen2010acoustic}.

For elastodynamics, however, it was found \cite{milton2006cloaking} that the equations change form under coordinate transformation and the cloaking ideas from electromagnetism and acoustics cannot be directly applied to elastic waves in solids. Specifically, the elastodynamic wave equations at a fixed frequency are given by:
\begin{eqnarray}\label{el}
\displaystyle \nonumber \nabla\cdot\boldsymbol{\sigma}=-\omega^2\rho\mathbf{u};\quad \boldsymbol{\sigma}=\mathbf{C}\nabla\mathbf{u}
\end{eqnarray}
which, under the coordinate transformation $\mathbf{x}'=\mathbf{x}'(\mathbf{x})$, transform to:
\begin{eqnarray}\label{elt}
\displaystyle \nabla '\cdot\boldsymbol{\sigma}'=\mathbf{D}'\nabla '\mathbf{u}'-\omega^2\boldsymbol{\rho}'\mathbf{u}';\quad \boldsymbol{\sigma}'=\mathbf{C}'\nabla '\mathbf{u}'+\mathbf{S}'\mathbf{u}'
\end{eqnarray}
The modified constitutive parameters are given by \cite{milton2006cloaking}:
\begin{eqnarray}\label{eltp}
\displaystyle \nonumber C^{'}_{pqrs}=\frac{1}{det\mathbf{A}}A_{pi}A_{qj}C_{ijkl}A_{rk}A_{sl};\quad S^{'}_{pqr}=\frac{1}{det\mathbf{A}}A_{pi}A_{qj}C_{ijkl}B_{rkl}\\
D^{'}_{pqr}=\frac{1}{det\mathbf{A}}B_{pij}C_{ijkl}A_{qk}A_{rl};\quad \rho^{'}_{pq}=\frac{1}{det\mathbf{A}}\left[\rho A_{pi}A_{qi}+B_{pij}C_{ijkl}B_{qkl}\right]
\end{eqnarray}
where $A_{ij}=\partial x^{'}_i/\partial x_j$ and $B_{ijk}=\partial^2x^{'}_{i}/\partial x_j\partial x_k$. This lack of invariance has made it hard to transfer the success of electromagnetic and acoustic cloaking to elastic wave cloaking. Moreover, the additional tensors $\mathbf{D}',\mathbf{S}'$ which appear in (\ref{elt}) are highly unusual as far as conventional materials are concerned. However, these parameters are not unusual for a Willis material. In fact, Eq. (\ref{homogenized}), under a fixed frequency is given by:
\begin{eqnarray}\label{homogenizedF}
\displaystyle \nonumber \langle\boldsymbol{\sigma}\rangle=\mathbf{C}^\mathrm{eff}\otimes\langle\boldsymbol{\varepsilon}\rangle-i\omega\mathbf{S}^\mathrm{eff}\otimes\langle{\mathbf{u}}\rangle\\
\langle\mathbf{p}\rangle=\bar{\mathbf{S}}^\mathrm{eff}\otimes\langle\boldsymbol{\varepsilon}\rangle-i\omega\boldsymbol{\rho}^\mathrm{eff}\otimes\langle{\mathbf{u}}\rangle
\end{eqnarray}
where $\otimes$ refers to convolution in space. With the following kinematic and dynamic relations:
\begin{eqnarray}
\displaystyle \langle\boldsymbol{\varepsilon}\rangle=\left[\nabla\langle\mathbf{u}\rangle+(\nabla\langle\mathbf{u})\rangle^\mathrm{T}\right]; \quad \nabla\cdot\langle\boldsymbol{\sigma}\rangle=-i\omega \langle \mathbf{p}\rangle
\end{eqnarray}
Eq. (\ref{homogenizedF}) transforms into the form of Eq. (\ref{elt}). Moreover, it can be shown that Eq. (\ref{elt}), under the coordinate transformation $\mathbf{x}''=\mathbf{x}''(\mathbf{x}')$, retains its form. Therefore, it appears that if elastic materials can be defined by the Willis form then the lessons learned from transformational optics and acoustics can be applied to elastic wave cloaking. It must be mentioned here that Willis material is not the only way of achieving elastic wave cloaking. Norris \cite{norris2011elastic} showed that Cosserat type materials \cite{cosserat1909theorie} would also serve the purpose. Some other techniques which have been studied for elastic wave cloaking include using anisotropy \cite{amirkhizi2010stress} and nonlinear prestressing \cite{parnell2012nonlinear,norris2012hyperelastic,parnell2012employing,parnell2013antiplane}.

\section{Conclusions}
In this paper we have approached the problem of acoustic/elastic metamaterials from two different directions, theory and applications. On the theoretical side we seek to define those effective dynamic homogenized relations which would describe the propagation characteristics of the mean wave in a composite. It is clear that the Willis constitutive relations (\ref{homogenized}) are the natural consequence of dynamic averaging and can be safely taken to apply to the homogenized dynamic behaviour of any elastic composite. Additionally, the Willis relations are found to be closed under homogenization whereas the conventional elastodynamic constitutive relations are not. On the application side we seek to control wave trajectories in certain specific ways and search for materials which would allow us to exert such control. Inevitably we find ourselves searching for materials with unusual properties which can, often, only be achieved in a homogenized sense. This is the common ground for theory and applications in acoustic/elastic metamaterials research. Some techniques (coordinate transformations) which have served well for the control of electromagnetic waves, when applied to acoustic/elastic waves, reinforce the idea that the Willis relations are a good description of inhomogeneous elastodynamics. This results from the fact that the Willis relations are closed under coordinate transformations as well. However, Willis relations are not unique and this nonuniqueness is mirrored in the practical manifestation that several different routes exist to achieve the aims of wave control in applications such as cloaking.

\section{Acknowledgments}:
The author acknowledges the support of the UCSD subaward UCSD/ONR W91CRB-10-1-0006 to the Illinois Institute of Technology (DARPA AFOSR Grant RDECOM W91CRB-10–1-0006 to
the University of California, San Diego).

%\bibliography{../References/ReferencesBib}

\end{document}